\documentclass[reprint,aps,prb,groupedaddress]{revtex4-1}

\usepackage{graphicx}

\begin{document}


\title{Infrared anomalous Hall effect in SrRuO$_3$: Evidence for crossover to intrinsic behavior}


\author{M.-H. Kim,$^1$ G. Acbas,$^1$ M.-H. Yang,$^1$ M. Eginligil,$^1$ P. Khalifah,$^2$ I. Ohkubo,$^3$
 H. Christen,$^4$ D. Mandrus,$^4$ Z. Fang,$^5$ and J. Cerne$^{1}$}

\affiliation{$^1$Department of Physics, University at Buffalo, The
State University of New York, Buffalo, NY 14260, USA}

\affiliation{$^2$Department of Chemistry, University at Stony
Brook, The State University of New York, Stony Brook, NY 11794,
USA}

\affiliation{$^3$Department of Applied Chemistry, University of
Tokyo, Tokyo, Japan}

\affiliation{$^4$Oak Ridge National Laboratory, Condensed Matter
Sciences Division, Oak Ridge, TN 37831, USA}

\affiliation{$^5$Institute of Physics, Chinese Academy of Science,
Beijing, 100080,China.}



\begin{abstract}
The origin of the Hall effect in many itinerant ferromagnets is
still not resolved, with an anomalous contribution from the sample
magnetization that can exhibit extrinsic \cite{smit, smit2, kats,
khalifah3} or intrinsic \cite{karplus, berger, ong, jungwirth,
fang} behavior.~\cite{AHE_review} We report the first mid-infared
(MIR) measurements of the complex Hall ($\theta_H$), Faraday
($\theta_F$), and Kerr ($\theta_K$) angles, as well as the Hall
conductivity ($\sigma_{xy}$) in a SrRuO$_3$ film in the
115-1400~meV energy range. The magnetic field, temperature, and
frequency dependence of the Hall effect is explored. The MIR
magneto-optical response shows very strong frequency dependence,
including sign changes. Below 200 meV, the MIR $\theta_H (T)$
changes sign between 120 and 150~K, as is observed in dc Hall
measurements. Above 200 meV, the temperature dependence of
$\theta_H$ is similar to that of the dc magnetization and the
measurements are in good agreement with predictions from a band
calculation for the intrinsic anomalous Hall effect
(AHE).~\cite{fang} The temperature and frequency dependence of the
measured Hall effect suggests that whereas the behavior above
200~meV is consistent with an intrinsic AHE, the extrinsic AHE
plays an important role in the lower energy response.
\end{abstract}

\pacs{}

\maketitle

\section{Introduction}

Ca$_x$Sr$_{1-x}$RuO$_3$ compounds exhibit unusual properties, such
as metamagnetism, quantum criticality, non-Fermi liquid behavior
and an anomalous Hall effect (AHE) that continue to challenge the
condensed matter community.  The Hall effect in SrRuO$_3$ consists
of two parts, the ordinary Hall effect (OHE) due to a magnetic
field $B$ producing a Lorentz force on moving carriers and the AHE
due to the sample's magnetization $M$. The Hall resistivity
$\rho_{H}$ is given by:
\begin{equation}
\rho_H = \rho_{yx} = R_0 B +
\rho_{yx}^{\mathrm{AHE}}(M)\label{eq;rhoH}
\end{equation}
The ordinary Hall coefficient $R_0$ is related to the carrier
density. The anomalous Hall resistivity
$\rho_{yx}^{\mathrm{AHE}}(M)$ can be divided into two categories:
1) the extrinsic AHE arising from impurity scattering and 2) the
intrinsic AHE due to the band structure.~\cite{AHE_review} The
anomalous Hall resistivity is expressed as
\begin{equation}
\rho_{yx}^{\mathrm{AHE}} (M) = R_s(\rho_{xx}) 4\pi M +
\sigma_{yx}^{I}(M) \rho_{xx}^2,\label{eq;AHE_rho}
\end{equation}
where $R_s$ is the extrinsic AHE coefficient as a function of the
longitudinal resistivity $\rho_{xx}$ and $\sigma_{yx}^{I}$ is the
intrinsic AHE transverse conductivity. The coefficient $R_s$
contains two terms, one proportional to $\rho_{xx}$ and one that
varies as $\rho_{xx}^2$:
\begin{equation}
R_s(\rho_{xx})=a \rho_{xx} + b \rho_{xx}^2, \label{eq;RS}
\end{equation}
where $a$ and $b$ are coefficients for two types of scattering
processes. The first term is due to asymmetric (skew) scattering
from impurities in the presence of magnetic order~\cite{smit,
smit2} and is considered to be a classical extrinsic effect. The
second term is called ``side-jump scattering" from impurities,
which has a quantum mechanical origin.~\cite{karplus, berger, ong}
Note that although the side-jump scattering scales as
$\rho_{xx}^2$ just as the intrinsic AHE, and some consider them to
be equivalent, the latter is simply proportional to $M$ while the
former depends on $M$ through the function $\sigma_{xy}^I (M)$.
While the dc AHE is observed in many itinerant ferromagnetic
materials ranging from ruthenates to colossal magnetoresistance
oxides to diluted magnetic semiconductors, the degree to which its
origin is intrinsic \cite{fang, lee2, jungwirth, ong} or extrinsic
\cite{smit, kats} is still not resolved in many
cases.~\cite{AHE_review} The dc AHE measurements by
Ref.~\onlinecite{kats} have found that the sign change of
$\theta_H (T)$ occurs in SrRuO$_3$ when $\rho_{xx} = 107~\mu
\Omega$~cm. They considered only the extrinsic AHE and found this
critical resistivity when $R_s = 0$, where the sign changes.

There have been extensive dc measurements~\cite{kats,khalifah3} to
probe the temperature and magnetic field dependence of the Hall
effect in SrRuO$_3$. There has also been a low temperature study
based on Kerr measurements to probe the frequency dependence of
the Hall conductivity above 200~meV.~\cite{fang} To the best of
our knowledge this work represents the first systematic study of
the temperature, magnetic field, and frequency dependence of the
Hall response in SrRuO$_3$ in the 0.1 - 1.4 eV energy range. This
range is particularly interesting since the only calculation to
model the frequency dependence of the Hall conductivity
$\sigma_{xy}$ of SrRuO$_3$ predicts strong spectral features for
$\sigma_{xy}$ in this range.~\cite{fang} This work provides the
first experimental test of this model below 200~meV.

Conventional dc Hall effect measurements in novel electronic
materials such as high temperature superconducting cuprates
(HTSC),~\cite{dchall} diluted nagnetic semiconductors, and
ruthenate perovskite  materials have been essential in revealing
the unusual character of these systems. Three major issues
motivate the study of the MIR AHE in these materials. The first
and most general argument is that since the understanding of dc
AHE may be unresolved, studying the frequency dependence of the
dynamic AHE can provide new information on the microscopic causes
that are responsible for the AHE. Secondly, MIR measurements
\cite{achall} would probe more effectively the energy scales of
the system (e.g., the plasma frequency, the cyclotron frequency
and the carrier relaxation rates) and provide greater insight into
the intrinsic electronic structure of a wide range of
materials.~\cite{au, cerne_ybco} The MIR range is highly
appropriate since the typical band energy scale for the AHE in
diluted magnetic semiconductors,~\cite{jungwirth, sinova,
Acbas_Kim_PRL09} SrRuO$_3$,~\cite{fang} and other materials~\cite{
broderick, sukhorukov} is in the MIR. It is not surprising that
the most striking spectral features in calculations of the AHE are
in the MIR range. Unlike conventional MIR spectroscopy that
measures the longitudinal conductivity $\sigma_{xx}$, which is
related to the \underline{sum} of the sample's response to left
and right circularly polarized, MIR Hall measurements probe the
Hall conductivity $\sigma_{xy}$, which is related to the
\underline{difference} in the sample's response to left and right
circularly polarized light, and therefore is more sensitive to
asymmetries such as spin-splitting etc. Finally, the dc Hall
effect can be dominated by impurity scattering or grain boundary
effects. This is especially important in new materials which often
contain many impurities and defects. By probing the Hall effect at
higher frequencies, the contribution from extrinsic scattering can
be minimized.

Magneto-polarimetry measurements can be used to extend Hall effect
measurements into the infrared frequency range ($10^{13}$~Hz).
These measurements are sensitive to the complex Faraday $\theta_F$
and Kerr $\theta_K$ angles, which are closely related to the
complex Hall angle $\theta_H$.~\cite{au} Since $\theta_H$ (and
$\theta_F$) obeys a sum rule~\cite{drew} it is very useful to be
able to integrate $\theta_H$ to higher frequencies to verify
whether (and where) the Hall angle sum rule saturates or whether
there is more relevant physics at even higher frequencies.
Finally, since the high frequency behavior of $\theta_H$ is
constrained by the general requirements of response functions, a
simple, model-independent asymptotic form for $\theta_H$ becomes
more accurate at higher frequencies.

In this paper, we report measurements of the MIR complex Faraday,
Kerr, and Hall angles in SrRuO$_3$ films in the 0.1 - 1.4 eV
energy range. The transmitted and reflected magneto-optical
responses in the MIR are qualitatively similar to results from dc
Hall and dc magnetization measurements. The measurements are in
good agreement with predictions from a band calculation
\cite{fang} above 200~meV. The deviations at lower energy are
probably due to a stronger contribution from the extrinsic AHE.

\section{Experimental System}

The Faraday and Kerr angles are measured using a sensitive
polarization modulation technique~\cite{instru_kim, verdet_kim} in
the mid and near-infrared (MNIR) spectral range (115 - 1400 meV)
for a SrRuO$_3$ sample grown by pulsed laser deposition at Oak
Ridge National Laboratory as described in
Ref.~\onlinecite{khalifah2}. Several light sources such as various
gas lasers, semiconductor lasers, and a custom-modified double
pass prism monochromator with a Xe light source allow us to
perform the measurement in a wide probe energy range. For details
of the experimental technique see Refs.~\onlinecite{instru_kim}
and~\onlinecite{verdet_kim}. The complex $\theta_F$ and $\theta_K$
angles are measured in the MNIR spectral range as a function of
magnetic field up to 2~T and temperature from 10~K to 300~K. The
small background Faraday signal from the cryostat windows and the
film substrate has been subtracted from the data. The complex
conductivities $\sigma_{xx}$ and $\sigma_{xy}$ and the complex
Hall angle $\theta_H$ are determined directly from the measured
complex $\theta_F$ and $\theta_K$ using the analysis techniques in
Ref.~\onlinecite{instru_kim}.

The dc longitudinal and Hall resistivities of SrRuO$_3$ were
measured simultaneously using a four-probe van der Pauw geometry
at UB's Magneto-Transport Facility as well as at Oak Ridge
National Laboratory using a Physical Property Measurement System
(PPMS, Quantum Design) in a six-terminal configuration and 1~mA
currents.~\cite{khalifah2} Moreover, the film magnetization was
measured by a SQUID magnetometer (Magnetic Property Measurement
System; MPMS, Quantum Design) using a custom-modified sample
holder, which enables measurements of the magnetization
perpendicular to the film surface, which is the same configuration
as our Faraday and Kerr measurements.

\section{Results}


The Faraday and Kerr angles on the SrRuO$_3$ were measured at
various temperatures and a wide energy range to probe the
anomalous Hall effect. The continuous broadband measurements allow
us to extend $\theta_F$ and $\theta_K$ measurements to up to 1.4
eV.~\cite{verdet_kim} Although the intensity of broadband light is
significantly weaker than that of lasers, and therefore the
sensitivity is reduced, $\theta_F$ and $\theta_K$ are large and
can be readily measured in this range. The analysis techniques to
obtain the complex longitudinal ($\sigma_{xx}$) and transverse
($\sigma_{xy}$) conductivities and the MIR Hall angles
($\theta_H$) are based on the techniques in
Ref.~\onlinecite{instru_kim}.


\begin{figure}[t]
\centering \includegraphics[scale=0.47, bb = 49 100 527 553, clip
= true]{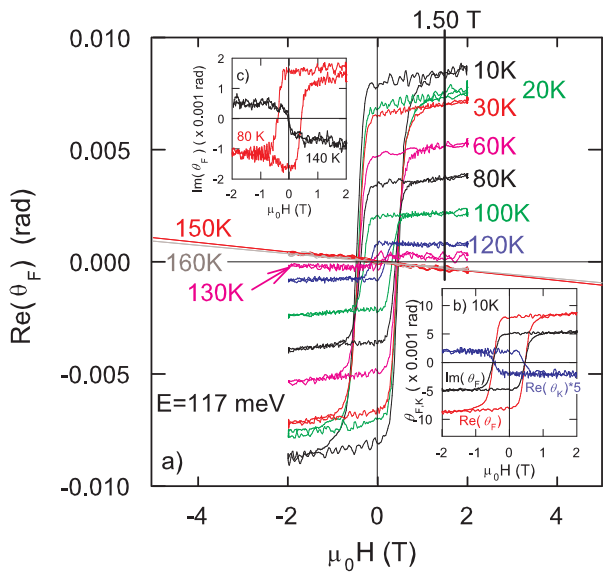} \caption{\label{fig;loops} Temperature dependence of
$\theta_F$ and $\theta_K$ from a SrRuO$_3$ as a function of
applied magnetic field $H$ at a probe energy of 117~meV. The main
panel a) shows the temperature dependence of the hysteresis loops
for $\mathrm{Re}(\theta_F)$. Linear fits to the data at 150 and
160~K are also shown to help distinguish the data from other
temperatures. The vertical line indicates $H$=1.5T, where the
hysteresis loops close at low temperatures. Inset b) shows
$\mathrm{Re}(\theta_F)$, $\mathrm{Im}(\theta_F)$, and
$\mathrm{Re}(\theta_K)$ (which is multiplied by a factor of 5) as
a function of $H$ at 117~meV and 10~K. Inset c) shows the
hysteresis loops for $\mathrm{Im}(\theta_F)$ at 117 meV and two
temperatures, 80 K and 140 K, at which the sign change near $H =
0$ of $\mathrm{Im}(\theta_F)$ is clearly seen.}
\end{figure}

Figure~\ref{fig;loops} shows temperature dependence of $\theta_F$
and $\theta_K$ in SrRuO$_3$ as a function of applied magnetic
field $H$ at a photon energy of 117~meV. Both $\theta_F$ and
$\theta_K$ exhibit ferromagnetic hysteresis until the
magnetization disappears at the Curie temperature $T_c \simeq
160$K. Figure~\ref{fig;loops}a) presents $\mathrm{Re}(\theta_F)$
as a function of $H$ at various temperatures. Note that the sign
of the slope above 130 K is opposite of that observed at lower
temperatures. Linear fits to the data at 150 and 160 K indicate
non-monotonic behavior of the slope. The strongest change in
$\mathrm{Re}(\theta_F)$ occurs below 1.5 T, where the
ferromagnetic hysteresis loops close.

The Figure~\ref{fig;loops}b) shows $\theta_F$ and $\theta_K$ as a
function of $H$ at 117 meV and 10 K. The magnitude of
$\mathrm{Im}(\theta_K)$, which is not shown in
Fig.~\ref{fig;loops}b), is an order of magnitude smaller than that
of $\mathrm{Re}(\theta_K)$. Typically, for metallic films like
SrRuO$_3$ in the MIR, the magneto-optic response in transmission
($\theta_F$) is larger than that obtained in reflection
($\theta_K$), although the transmitted intensity of light can be
as small as 0.01~\%, because $\theta_F \propto
\sigma_{xy}/\sigma_{xx}$ while $\theta_K \propto
\sigma_{xy}/(\sigma_{xx})^2$ where $\sigma_{xx} \gg \sigma_{xy}$
in metallic films.~\cite{instru_kim} For thicker and more metallic
films, $\theta_F$ is much more sensitive to $\sigma_{xy}$ than
$\theta_K$ in the MIR. Therefore, measuring magneto-optic signals
in both transmission and reflection with a higher sensitivity
provides the measurements reported here with a distinct advantage
over the technique used in Refs.~\onlinecite{fang} and
\onlinecite{fang_som}, where only the reflected signal was
measured. Moreover, the stronger $\theta_F$ signals are critical
to explore the MIR Hall effect at higher temperatures, where the
magneto-optical response is weaker.


\begin{figure}[t]
\centering \includegraphics[scale=0.47, bb = 37 150 487 718, clip
= true]{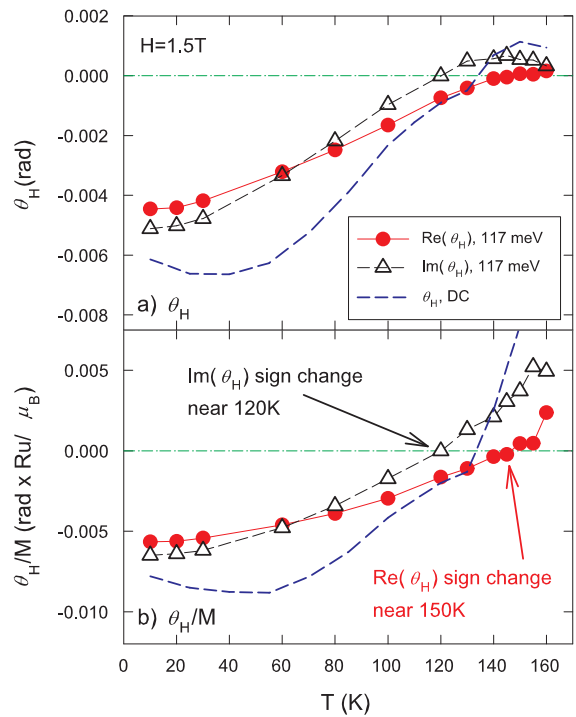} \caption{\label{fig;Hall_T} Temperature dependence
of a) $\theta_H (T)$ and b) $\theta_H (T) / M(T)$ at 1.5~T at
117~meV and dc. The magnetization $M(T)$ is shown in
Fig.~\ref{fig;Hall_T_highE}.}
\end{figure}

It is not clear whether the sign change in $\mathrm{Re}(\theta_F)$
is due to a sign change in the AHE component or the OHE component
with opposite sign (from opposite charge carriers) which starts
dominating at higher temperatures in Fig.~\ref{fig;loops}a).
However, the AHE nature of the sign change is clearly seen in the
Fig.~\ref{fig;loops}c), which shows $\mathrm{Im}(\theta_F)$ at 80
K and 140 K at a probe energy of 117 meV. Here the sharp step near
$H \simeq 0$ changes sign. The response of $\mathrm{Re}(\theta_F)$
is quite linear in $H$ at the sign change, which occurs in
temperatures between 130 K and 160 K where the magnetization is
smaller due to the proximity to $T_c$. On the other hand, unlike
$\mathrm{Re}(\theta_F)$, the sign change in
$\mathrm{Im}(\theta_F)$ appears between 80 K and 140 K far $T_c$.
The opposite signs for the hysteresis loop steps in
$\mathrm{Im}(\theta_F)$ indicate the sign change is related to the
magnetization, and hence due to the sign change in the AHE
coefficient $R_s$. The sign change in the ferromagnetic step near
$H = 0$ is not clearly seen in dc Hall measurements, which
demonstrates another advantage of magneto-optical measurements in
probing AHE materials.

Figure~\ref{fig;Hall_T} shows the Hall angle as a function of
temperature at an applied magnetic field of 1.5~T and at a probe
energy of 117 meV and 0 meV (dc). Since the hysteresis loops close
below $\mu_0 |H|$=1.5~T, this value of magnetic field is chosen to
characterize the strength of the magneto-optical response. It is
difficult to separate the OHE from the AHE above 1.5 T, because
both Hall signals are linear in $H$. Reference~\onlinecite{kitama}
has reported that the magnetization in SrRuO$_3$ does not saturate
even at magnetic fields of 40~T and found that the anomalous Hall
signal can be linear in $H$ up to 40~T. Although the separation of
AHE from OHE has been done in dc measurements,~\cite{kats} it is
more challenging to make this separation for MIR data. However,
there are two regions where the OHE signal can be readily
separated from the AHE signal. First of all, at lower temperatures
and low magnetic fields, where magneto-optical response is
ferromagnetic, $\theta_H$ is dominated by the AHE. For higher
temperatures, it is still hard to separate AHE from OHE because
the ferromagnetic response is weaker as the temperature increases
towards $T_c$. Second of all, the OHE is suppressed at higher
frequency. The Drude model predicts that at as the probe frequency
$\omega$ increases past the characteristic scattering frequency
$\gamma_H$, $\mathrm{Re}(\theta_H)\propto \omega^{-2}$ and
$\mathrm{Im}(\theta_H)\propto \omega^{-1}$.~\cite{au} If one
assumes that the linear part of the Hall angle ($\propto
\theta_F$) only comes from the OHE due to free carriers in
Fig.~\ref{fig;loops}, one can use the extended Drude model to
calculate the Hall frequency $\omega_H$, which is related to the
carriers' effective mass, and the Hall scattering rate
$\gamma_H$.~\cite{au,cerne_ybco} Furthermore, one can estimate the
maximum contribution of the OHE to the overall Hall response.
$\theta_F$ and $\theta_K$ produce $\sigma_{xx}$, $\sigma_{xy}$ and
$\theta_H = \sigma_{xy} / \sigma_{xx}$. Slope of the linear
behavior of the $\theta_F$ (and therefore $\theta_H$) in
Fig.~\ref{fig;loops} at fields above 1.5~T is nearly constant from
10 K to 80 K. The slope translates into $\omega_H = -0.31$
cm$^{-1}$/T ($-0.039$ meV/T) and $\gamma_H = 1102$ cm$^{-1}$ (137
meV). Assuming that the carrier scattering is isotropic, one can
use $\omega_H$ to obtain a carrier effective mass of $2.99 m_e$,
where $m_e$ is the bare electron mass. Since the Hall scattering
rate $\gamma_H$ is close to $\omega =$ 117 meV, one can expects
that $\mathrm{Re}(\theta_H) \approx \mathrm{Im}(\theta_H) \approx
\omega_H / (2\gamma_H)$ when $\omega \approx \gamma_H$. The value
of OHE given by the linear part of the hysteresis is less than 3\%
of the total Hall response. Likewise, the OHE is much smaller at
the higher frequencies over 117 meV due to $\omega \gg \gamma_H$.

\begin{figure}[t]
\centering \includegraphics[scale=0.47, bb = 38 109 558 730, clip
=  true]{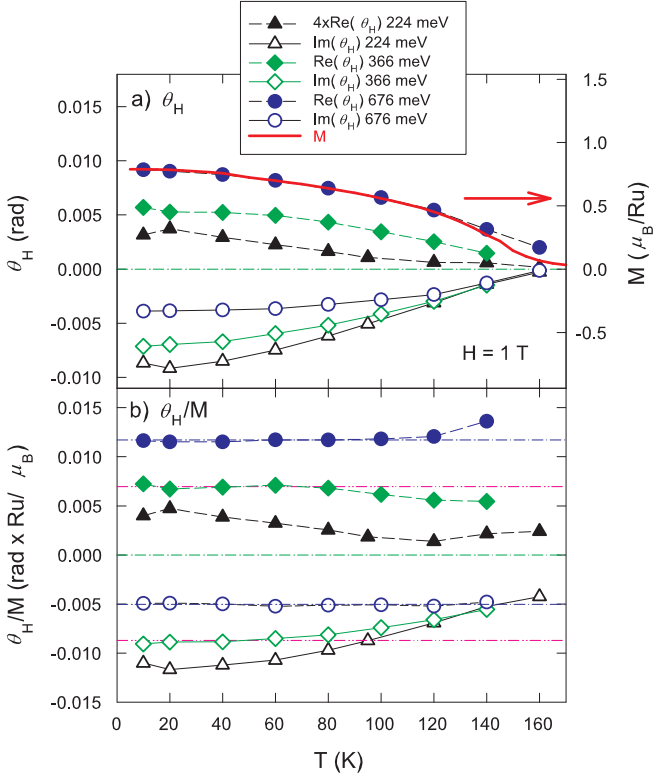} \caption{\label{fig;Hall_T_highE} Temperature
dependence of a) $\theta_H (T)$ and $M(T)$ and b) $\theta_H (T) /
M(T)$ at 1~T above 200~meV. $\theta_H(T)$ is very similar to the
magnetization $M(T)$ at 1~T. The horizontal lines in b) are guide
lines, which show $\theta_H(T) \propto M(T)$. }
\end{figure}

Figure~\ref{fig;Hall_T}a) shows the sign change of
$\mathrm{Re}(\theta_H)$ and $\mathrm{Im}(\theta_H)$ at 117 meV
near 130 K, where the Hall sign change is observed in dc
measurements (dashed line). The magnitude and temperature behavior
of $\theta_H$ at 117 meV and dc are very similar. The sign change
of $\mathrm{Im}(\theta_H)$ appears at a slightly lower
temperature, deeper in the ferromagnetic phase, than
$\mathrm{Re}(\theta_H)$. It may explain why the sign change of
$\mathrm{Im}(\theta_H)$ is more clearly seen in
Fig.~\ref{fig;loops}. The magnitudes of both
$\mathrm{Re}(\theta_H)$ and $\mathrm{Im}(\theta_H)$ are slightly
smaller than that of dc. The dc Hall angle exhibits a minimum
value near 40 K, but at 117 meV both $\mathrm{Re}(\theta_H)$ and
$\mathrm{Im}(\theta_H)$ increase monotonically with temperature.
$\mathrm{Im}(\theta_H)$ at 117 meV rises more steeply than
$\mathrm{Re}(\theta_H)$, changes sign earlier, and reaches a
maximum value near 140 K. Since the MIR Hall angles mostly come
from the AHE, the $\theta_H$ is associated with the magnetization
$M$. Figure~\ref{fig;Hall_T}b) shows the Hall angles divided by
the measured $M$, which is shown in Fig.~\ref{fig;Hall_T_highE}a)
(solid line). The sign change in $\mathrm{Re}(\theta_H)/M$ and
$\mathrm{Im}(\theta_H)/M$ is clearly seen. The sign change in
$\mathrm{Im}(\theta_H)$ occurs near 120 K and in
$\mathrm{Re}(\theta_H)$ near 150 K. These are symmetrically
separated from the dc Hall sign change temperature.


Figure~\ref{fig;Hall_T_highE} shows the temperature dependence of
the dc magnetization and $\theta_H$ at higher energies of 224,
366, and 676~meV, which is obtained from the formula of $[\theta_H
(+1T) - \theta_H (-1T)]/2$. In this case, $\theta_H (T)$ no longer
changes sign as it does at 117 meV and dc measurements, instead
$\theta_H (T)$ approaches zero gradually as shown in
Fig.~\ref{fig;Hall_T_highE}a). As the probe energy increases,
$\theta_H (T)$ behaves more like the magnetization $M$. The solid
line in Fig.~\ref{fig;Hall_T_highE}a) represents the measured
magnetization, which is saturated at lower temperature to 0.8
$\mu_\mathrm{B} / \mathrm{Ru}$. At an energy of 676 meV, $M(T)$
and $\theta_H (T)$ exhibit the same dependence over nearly the
entire range as shown in Fig~\ref{fig;Hall_T_highE}a). Typically,
visible and near-infrared Faraday and Kerr measurements are used
to determine the magnetization of materials.
Figure~\ref{fig;Hall_T_highE}b) plots $\theta_H (T) / M(T)$ at
higher energies of 224, 366, and 676~meV at 1~T. The horizontal
guide lines show behavior where $\theta_H$ is proportional to $M$.
At a probe energy 676 meV the value of $\theta_H / M$ is a
constant at all temperatures, but at 366 meV this is the case only
at lower temperatures.

\begin{figure}[t]
\centering \includegraphics[scale=0.47, bb = 55 103 513 670, clip
= true]{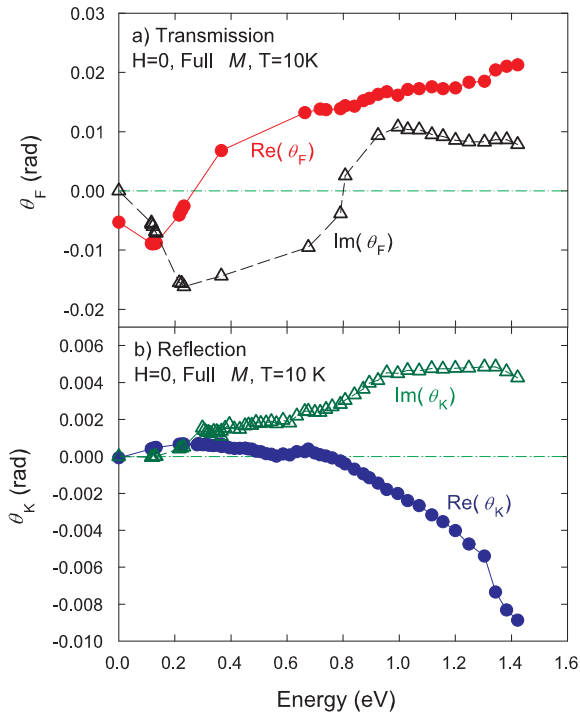} \caption{\label{fig;far_kerr} Energy dependence of
the AHE a) $\theta_F$ and b) $\theta_K$ with the sample fully
magnetized perpendicular to the plane at 0~T and 10~K. Since the
measurement is at $H=0$~T, the OHE as well as background signals
from the substrate and windows, which are linear in $H$, do not
contribute to the signal. At the energy range from 0.4 - 0.7 eV,
the intensity of transmitted light is so weak to measure
$\theta_F$.}
\end{figure}

Figure~\ref{fig;far_kerr} shows the anomalous Hall part of
$\theta_F$ and $\theta_K$ at 10 K and 0 T with the sample fully
magnetized out of plane as a function of probe energy. As seen
before, the OHE contribution is small, especially at higher
energies at $H \approx 1$~T. In addition, since the measurement is
performed at $H$ = 0~T and OHE is proportional to $H$, even small
OHE cannot contribute to this Hall signal. Both $\theta_F$ and
$\theta_K$ display a strong energy dependence and change sign
mostly at lower energies, but monotonically increase or remain
fairly constant at higher energies as shown in
Fig.~\ref{fig;far_kerr}. The energy dependence of
$\mathrm{Re}(\theta_F)$ and $\mathrm{Im}(\theta_K)$ is similar.
The sign changes are observed at low energies near 250 meV for
$\mathrm{Re}(\theta_F)$ and 130 meV for $\mathrm{Im}(\theta_K)$.
Both signs change from negative to positive as energy increases.
Note that the signs are defined and determined in
Ref.~\onlinecite{instru_kim}. Likewise, $\mathrm{Im}(\theta_F)$
and $\mathrm{Re}(\theta_K)$ exhibit similar features. However,
both sign changes appear near 800 meV. The sign of
$\mathrm{Im}(\theta_F)$ also changes from negative to positive,
but the sign of $\mathrm{Re}(\theta_K)$ behaves oppositely. The dc
values of $\theta_F$ and $\theta_K$ in Fig.~\ref{fig;far_kerr} are
determined in Ref.~\onlinecite{instru_kim} connect smoothly with
the MIR data as $E \rightarrow 0$.

\begin{figure}[t]
\centering \includegraphics[scale=0.47, bb = 51 103 513 671, clip
= true]{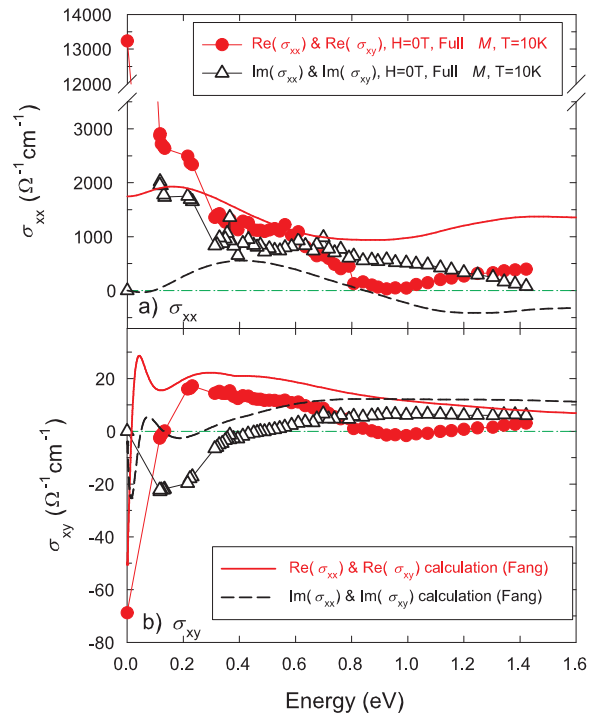} \caption{\label{fig;sigma} a) The longitudinal
conductivity $\sigma_{xx}$ and b) transverse (AHE) conductivity
$\sigma_{xy}$ b) as a function of probe energy. The thin lines are
from a Berry phase calculation of the intrinsic AHE by Z. Fang.
Note that this calculation neglects intraband transitions.}
\end{figure}

Figure~\ref{fig;sigma} shows the measured and calculated complex
a) longitudinal conductivity $\sigma_{xx}$ and b) transverse (AHE)
conductivity $\sigma_{xy}$ as a function of probe energy. The
symbols are obtained from $\theta_F$ and $\theta_K$ in
Fig.~\ref{fig;far_kerr} using the analysis techniques in
Ref.~\onlinecite{instru_kim}. One advantage of determining both
$\sigma_{xx}$ and $\sigma_{xy}$ from the same set of $\theta_F$
and $\theta_K$ measurements is that the behavior of $\sigma_{xx}$,
which is fairly well known in this energy range,~\cite{kostic} can
be provide a consistency test for $\sigma_{xy}$, which is not well
known. As experimentally seen in Ref.~\onlinecite{instru_kim}, the
complex $\sigma_{xx}$ from $\theta_F$ and $\theta_K$ is in good
agreement over the entire energy range with $\sigma_{xx}$ obtained
from reflectance measurements on a different SrRuO$_3$
film,~\cite{kostic} which has a factor of 3 smaller dc
resistivity. On the other hand, theoretical predictions for the
AHE have been limited at finite energies, although several
different models are used to explain the dc AHE in
SrRuO$_3$.~\cite{smit, berger, fang, kats} Here we compare the
measured MIR conductivities with predictions from a Berry-phase
calculation for SrRuO$_3$.~\cite{fang} The solid lines (the real
part of $\sigma_{xx}$ and $\sigma_{xy}$) and dashed lines (the
imaginary part of $\sigma_{xx}$ and $\sigma_{xy}$) are from a
Berry phase calculation of the intrinsic AHE by Z. Fang and
coworkers.~\cite{fang}  The calculated complex $\sigma_{xx}$ in
Fig.~\ref{fig;sigma}a) clearly deviates from the measured
$\sigma_{xx}$. This is not surprising since the model neglects
intraband transitions, which play a dominant role at lower
energies in a metal. For energies over 200~meV, the calculation
agrees qualitatively and in some energy regions quantitatively
with the measured value. So unlike the calculation, there is no
sign change of the measured $\mathrm{Im} (\sigma_{xx})$ in our
energy range.

Since the goal of this calculation was to provide insights into
the behavior of $\sigma_{xy}$, it is encouraging that the
agreement between the calculated and measured values of
$\sigma_{xy}$ is significantly better, as can be seen in
Fig.~\ref{fig;sigma}b), which shows the anomalous Hall response.
The Fang model predicts a sign change from electron-like at low
energy to hole-like at higher energies in both
$\mathrm{Re}(\sigma_{xy})$ and $\mathrm{Im}(\sigma_{xy})$. The
measured $\mathrm{Re}(\sigma_{xy})$ changes from electron-like for
energies below 150 meV to hole-like at higher energies. The
measured $\mathrm{Im}(\sigma_{xy})$ appears to make the same
change above 400 meV. Both $\mathrm{Re}(\sigma_{xy})$ and
$\mathrm{Im}(\sigma_{xy})$ extrapolate smoothly to their dc
values. Unlike conventional metals like gold, which can be modeled
using a single band in the infrared and where Drude behavior
(intraband transition) are responsible for the infrared Hall
effect, in SrRuO$_3$ both intraband and interband transitions
contribute, so the ``electron-like" description is only meant to
indicate signs, not the microscopic origin of the Hall effect. Of
course, in the dc limit the OHE is solely due to intraband (Drude)
behavior. Above 300 meV, the calculated $\mathrm{Re}(\sigma_{xy})$
and $\mathrm{Im}(\sigma_{xy})$ values run roughly parallel to the
measured ones. The crossing of $\mathrm{Re}(\sigma_{xy})$ and
$\mathrm{Im}(\sigma_{xy})$ is observed in both the calculated and
measured values near 1 eV and 0.8 eV, respectively. Below 300 meV,
$\mathrm{Re}(\sigma_{xy})$ exhibits a peak near 200 meV in both
the calculation and measurements. Another sharp peak-like
structure is predicted in $\mathrm{Re}(\sigma_{xy})$ near 50 meV.
However, the measured $\mathrm{Re}(\sigma_{xy})$ has already
dipped below zero at 120 meV, in strong contrast to the
theoretical upturn at 100 meV, and is heading towards the dc value
smoothly. For $\mathrm{Im}(\sigma_{xy})$, the calculation and the
measurements shows the zero-crossing near 300 meV and 500 meV,
respectively. As with the calculated $\mathrm{Re}(\sigma_{xy})$,
the calculated $\mathrm{Im}(\sigma_{xy})$ also exhibits sharp
peaks near 130 and 15 meV, but the slope and the large negative
offset of the measured $\mathrm{Im}(\sigma_{xy})$ suggests that it
will continue monotonically towards its dc value of zero. It is
interesting to note that the over the entire measured energy
range, the calculated $\sigma_{xy}$ generally agrees better with
the data than the calculated $\sigma_{xx}$. These measurements
strongly support the validity of the Berry phase model for
describing the anomalous Hall response of SrRuO$_3$ above 200 meV,
but below 200 meV the measurements do not follow theoretical
predictions.


\begin{figure}[t]
\centering \includegraphics[scale=0.47, bb = 47 94 510 662, clip =
true]{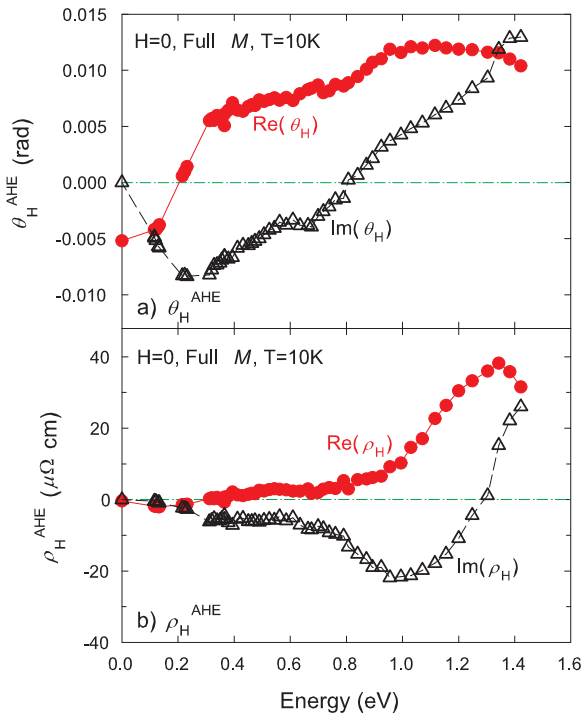} \caption{\label{fig;Hall_E} The complex a) Hall angle
$\theta_H$ and b) Hall resistivity $\rho_H$ at 10~K and and 0~T
with the sample fully magnetized out of plane as a function of
probe energy. Since the measurement is at $H=0$~T, the signals are
solely due to the AHE.}
\end{figure}

Figure~\ref{fig;Hall_E} shows the measured complex a) Hall angle
$\theta_H$ and b) Hall resistivity $\rho_H$ at 10 K and 0 T
(sample fully magnetized out of plane). For low energies,
$\mathrm{Re}(\theta_H)$ is nearly constant in the 0 - 120 meV
range, whereas $\mathrm{Im}(\theta_H)$ increases linearly in the
same range. The low energy MIR $\theta_H$ results extrapolate
smoothly to the dc values, suggesting that there are no additional
features in the 0 - 120 meV range. Above 100 meV,
$\mathrm{Re}(\theta_H)$ increases more rapidly with increasing
energy, changes sign near 200 meV, and increases monotonically
over most of measurement range. $\mathrm{Im}(\theta_H)$ reaches a
minimum value near 200 meV, changes sign near 800 meV, and becomes
more positive as energy increases. The energy dependence of
$\rho_H$ in Fig.~\ref{fig;Hall_E}b) is nearly constant below 0.8
eV, but above this energy the value of $\rho_H$ increases an order
of magnitude. Above 0.8 eV $\mathrm{Re}(\rho_H)$ has peak near 1.3
eV, where $\mathrm{Im}(\rho_H)$ changes sign. Additionally,
$\mathrm{Im}(\rho_H)$ has a minimum value near 1 eV.

\section{Discussion}

Exploring the Hall effect in SrRuO$_3$ as a function of magnetic
field $H$, temperature $T$, and frequency $\omega$ (or MNIR
energy) can provide new insights into the material as well as the
AHE in general. We have seen the MNIR $\theta_H$ (or
$\sigma_{xy}$) is consistent with the predictions from an
intrinsic AHE calculation for energy greater than 200 meV. On the
other hand, the frequency, temperature and magnetic field
dependence below 200 meV may be more consistent with expectations
from the extrinsic AHE. In this discussion, we explore the
evidence for a crossover between intrinsic and extrinsic behaviors
in the MIR energy range. Since the extended Drude model (EDM) was
successfully applied to the IR $\sigma_{xx}$ in
SrRuO$_3$~\cite{kostic}, we will compare EDM predictions for
$\sigma_{xx}$ and $\sigma_{xy}$ with our measurements in order to
disentangle OHE and AHE contributions. Secondly, we apply the
extrinsic AHE model to the temperature dependent $\theta_H$ at 117
meV to test this model at and near dc.

The EDM is more important at lower energy, because it models the
intraband transitions which are dominant at and near dc. Note that
the EDM considers the frequency dependent relaxation time
$\tau^\ast (\omega)$ and plasma frequency $\omega_p^\ast
(\omega)$, which are renormalized by the mass enhancement factor
$m^\ast (\omega)$. The longitudinal conductivity $\sigma_{xx}$ is
defined in EDM as
\begin{equation}
\label{eq;ex_Drude} \sigma_{xx} = \frac{\omega_p^\ast (\omega)^2}{
4\pi \left(\gamma^\ast (\omega) - i \omega \right)},
\end{equation}
where renormalized scattering rate is $\gamma^\ast (\omega) =
\tau^{\ast -1} (\omega) = \gamma (\omega)/ m^\ast (\omega)$, and
frequency-dependent plasma frequency $\omega_p^\ast (\omega)^2 =
\omega_p^2 / m^\ast (\omega)$. The bare plasma frequency
$\omega_p$ of the free electron gas can be calculated by the
carrier density. This model was used in SrRuO$_3$ to probe
possible non-Fermi-liquid behavior exhibited in reflectance
measurements.~\cite{kostic} It was reported that
$\mathrm{Re}(\sigma_{xx})$ falls like $\omega^{-0.5}$ and
$\gamma^\ast (\omega)$ increases linearly with $\omega$ in the
energy range between 0 to 1000 cm$^{-1}$ (124 meV) at low
temperature. The renormalized scattering rate is given by the
formula $\gamma^\ast (\omega) = \omega \left[
\mathrm{Re}(\sigma_{xx})/\mathrm{Im}(\sigma_{xx}) \right]$, which
is derived from Eq.~(\ref{eq;ex_Drude}), and using data in
Fig.~\ref{fig;Hall_E}a). The scattering rate $\gamma^\ast
(\omega)$ obtained from our $\sigma_{xx}$ increases fairly
linearly with $\omega$ in the energy range of 116 - 600 meV (900 -
4800 cm$^{-1}$) and 10 K. Furthermore, in the same energy range
the conductivity $\mathrm{Re}(\sigma_{xx})$ drops like
$\omega^{-0.68}$. Our results for $\sigma_{xx}$ based on
$\theta_F$ and $\theta_K$ measurements are consistent with that in
Ref.~\onlinecite{kostic}.

The EDM allows one to extract the carrier density $n$ of the
electron gas. One way to obtain $n$ is from the OHE. The Hall
scattering parameters are extracted from the linear in $H$
behavior of $\theta_H$ at low temperatures, assuming that this
slope is solely due to the OHE. At 117 meV we obtain $\omega_H =
-0.31 ~\mathrm{cm}^{-1}/\mathrm{T}$, $\gamma_H = 1102
~\mathrm{cm}^{-1}$, and a carrier effective mass of 2.99 $m_e$.
Likewise, the MIR Hall coefficient $R_H$ can be determined from
the linear behavior of $\theta_H (H)$. The coefficient $R_H$ is
given by:
\begin{equation}
R_H (\omega) = \frac{\rho_{yx}(\omega)}{B}=\frac{\theta_H
(\omega)}{\sigma_{xx}(\omega) B}, \label{eq;RH}
\end{equation}
In a simple Drude model where $\gamma$ is constant, $R_H (\omega)$
is purely real and constant. Considering only $\mathrm{Re}(R_H)$
at 117 meV, one can estimate $n = 1.04 \times 10^{22}$
electrons/cm$^3$ from using the formula $R_H = - \frac{1}{ne}$,
where $e$ is the electron charge. This value is nearly within a
factor of two of that obtained in dc measurements by Khalifah and
coworkers,~\cite{khalifah2}  where $n = 2.5\times
10^{22}$~electrons/cm$^3$ (or $n=1.6 \times
10^{22}$~electrons/cm$^3$ from Ref.~\onlinecite{izumi}). This
discrepancy is probably due to the fact that the linear behavior
is not completely due to the OHE, but also the AHE. If 50\% of the
net linear Hall angle signal at 117~meV between 1.5 and 2~T
resulted from the AHE, leading to a factor of two decrease in the
contribution from the OHE, the carrier density obtained from the
slope of $\theta_H$ at 117 meV, at 10 K, and above 1.5~T would be
$2.1 \times 10^{22}$~electrons/cm$^3$. The carrier density $n =
1.04 \times 10^{22}$~electrons/cm$^3$ corresponds to the plasma
frequency $\omega_p = 30517$~cm$^{-1}$, while $n = 2.5 \times
10^{22}$~electrons/cm$^3$ corresponds to $\omega_p =
47315$~cm$^{-1}$. According to Ref.~\onlinecite{kostic}, the mass
enhancement factor $m^\ast (\omega)$ approaches 1 as the energy
exceeds 1000 cm$^{-1}$. It means that $\omega_p^\ast (\omega)^2
\approx \omega_p^2$ is satisfied from 1000 cm$^{-1}$ up to 4800
cm$^{-1}$, where $\gamma^\ast (\omega) \propto \omega$. For the
EDM $\sigma_{xx}$, the appropriate partial sum is
\begin{equation}
\label{eq;sum_rule} \frac{\omega_p^\ast (\omega_c)^2}{4 \pi} =
\int_0^{\omega_c} \frac{2}{\pi} \mathrm{Re} \left[ \sigma_{xx}
(\omega) \right] d\omega,
\end{equation}
where $\omega_c$ is a cutoff frequency. To stay consistent with
the analysis in Ref.~\onlinecite{kostic}, we fit the measured
$\mathrm{Re}(\sigma_{xx})$ to the formula $A \omega^{-0.5}$, where
$A$ is a fitting parameter. If the partial sum is performed to
$\omega_c = 1000$~cm$^{-1}$, then $\omega_p^\ast =
32636$~cm$^{-1}$, which corresponds to a carrier density $n = 1.2
\times 10^{22}$~electrons/cm$^3$. For $\omega_c = 4800$~cm$^{-1}$,
$\omega_p^\ast = 48307$~cm$^{-1}$, which corresponds to $n = 2.6
\times 10^{22}$~electrons/cm$^{3}$. These values of carrier
density are qualitatively in agreement with the values obtained
from the OHE and dc measurements. The scattering parameters
determined from the EDM reveal that intraband transitions play an
important role in $\sigma_{xx}$ below 100 meV.

As with the MIR $\sigma_{xy}(T)$, the behavior of $\theta_H$ can
be divided into two distinct regions at 200 meV. Below 200 meV,
the MIR $\theta_H (T)$ is similar to the dc $\theta_H (T)$ as in
Figs.~\ref{fig;Hall_T} and~\ref{fig;Hall_T_highE}, but above 200
meV it looks like the dc magnetization $M(T)$. The agreement in
$\sigma_{xy}(\omega)$ between the measurement and the strictly
interband, intrinsic, Berry phase model at energies below 200 meV
is not as good as at higher energies as shown in
Fig.~\ref{fig;sigma}b). Moreover, the EDM analysis suggests that
intraband transitions are important to $\sigma_{xy}$ and even more
important to $\sigma_{xx}$ at energies below 100 meV. One possible
reason for discrepancy between measurement and theory at low
energies is that the lifetime broadening used in the calculation
is too small for the SrRuO$_3$ film measured here, which exhibits
much broader features. Another possibility is that the intrinsic
model does not include intraband transitions, which play an
important role near dc.

Since much of the dc AHE is commonly framed using
Eqs.~(\ref{eq;rhoH}) and (\ref{eq;RS}), it is interesting to study
the temperature and frequency dependence of $\theta_H$ in terms of
these equations. The simplest generalization to finite frequency
of the extrinsic AHE $\theta_H$ is to simply use the frequency
dependent resistivity $\rho_{xx}(\omega)$ in Eqs.~(\ref{eq;rhoH})
and (\ref{eq;RS}) at $H$ = $B$ = 0 T:
\begin{equation}
\theta_H (\omega) = \frac{\rho_H(\omega)}{\rho_{xx}(\omega)} =
\frac{R_s(\omega)}{\rho_{xx}(\omega)} 4 \pi M = ( a + b
\rho_{xx}(\omega) ) 4 \pi M. \label{eq;rhoH_E}
\end{equation}
Although this generalization is not rigorously correct, it can
provide insight into the lower energy AHE. If one considers only
the extrinsic AHE at or near dc, the sign change appears to be due
to a change in sign of the AHE coefficient $R_s$. In
Ref.~\onlinecite{kats}, the dc AHE always vanishes at the same
value of $\rho_{xx}^\ast = 107~\mu \Omega$ cm, where $\theta_H
(\rho_{xx}^\ast) = 0$, which suggests that the extrinsic AHE is
dominant at dc. In our dc measurements, the sign change occurs at
approximately $\rho_{xx}^\ast = 173~\mu \Omega$ cm. If one applies
Eq.~(\ref{eq;rhoH_E}) using $\rho_{xx}(\omega)$ to the sign change
at 117 meV, the sign change of $\mathrm{Re}(\theta_H)$ appears at
$\rho_{xx}^\ast = 400~\mu \Omega$~cm and that of
$\mathrm{Im}(\theta_H)$ occurs at $\rho_{xx}^\ast = 383~\mu
\Omega$ cm. Typically, the longitudinal resistivity at 117 meV is
two or three times larger than the value at dc because $\rho_{xx}
\propto \gamma (\omega) \propto \omega$ at lower energies.
According to Eq.~(\ref{eq;rhoH_E}), the sign change occurs when
$\rho_{xx}^\ast = - a/b$. The parameters $a$ (skew scattering) and
$b$ (side-jump scattering) are derived from microscopic
origins.~\cite{smit,smit2,berger} The analysis at 117 meV reveals
that $a$ and $b$ are energy-dependent. Therefore, from
$\rho_{xx}^\ast (\omega) = - a(\omega)/b(\omega)$ and since
$\rho_{xx}^\ast$ grows with $\omega$, it could be interpreted that
the skew scattering term $a(\omega)$ grows with respect to the
side-jump scattering term $b(\omega)$ as energy increases for 0 to
117 meV. Another point of view is that the skew scattering is more
sensitive to the probe energy than the side-jump scattering below
117 meV. Since the OHE is zero at $H = 0$~T  or negligibly small
even at $H \approx 0$ at the probe energy used here, the measured
value is mostly contributed by the AHE.

To sum up the results, the intrinsic AHE model qualitatively
agrees with the measurements above 200~meV and the disagreement
between the intrinsic AHE model and the measurements at low
energies is most likely due to the extrinsic AHE. This is a
reasonable conclusion, since the extrinsic AHE originates from
carrier scattering from impurities/defects. At low frequency, the
carriers scatter many times in each driving cycle, resulting in a
response that is similar to the dc Hall effect. At high driving
frequencies, the carriers oscillate many times before experiencing
a scattering event, and hence the Hall response is dominated by
the dynamics of the oscillations (band structure) and not by the
magnetic scattering. One could expect that the characteristic
frequency at which the extrinsic Hall effect begins to disappear
is the impurity scattering frequency. Since the total scattering
rate includes the impurity scattering rate, the extrinsic AHE
should start decreasing at a frequency at or below the total
scattering rate. It is interesting that the upper limit for the
impurity scattering rate is similar in magnitude as the frequency
above which: 1) the intrinsic AHE model begins to approach the
measurements of $\sigma_{xy}$; 2) the temperature dependence of
the MIR Hall effect loses its dc character; and 3) the Drude-like
behavior in $\sigma_{xx}$ significantly decreases. Therefore, the
upper limit for the impurity scattering rate appears to be near
1600~cm$^{-1}$ (200 meV). This confirms the importance of the
extrinsic AHE in dc and low frequcny measurements,~\cite{kats,
khalifah3} but also shows the transition to more intrinsic
behavior at higher frequencies. Our measurements suggest that by
increasing the probe energy beyond 200 meV, the AHE makes a
transition from an extrinsic to an intrinsic character. We find
that the Berry phase model accurately describes the MIR AHE of
SrRuO$_3$ above 200 meV.

\begin{acknowledgments}

We thank K. Takahashi, A.J. Millis, N.P. Ong, and J. Sinova for
helpful discussions. We also wish to thank B.D. McCombe for the
use of UB's Magneto-Transport Facility. This work was supported by
Research Corporation Cottrell Scholar Award,
NSF-CAREER-DMR0449899, and an instrumentation award from the
University at Buffalo, College of Arts and Sciences.

\end{acknowledgments}

\end{document}